\documentclass[12pt]{iopart}
\newcommand{\R}{{\mathbb{R}}}

\usepackage{epsfig,epsf,amsfonts,amscd}

\begin{document}
%\sloppy
%\draft
\title{ Families of one-point interactions resulting from the squeezing limit of the sum of 
two- and three-delta-like potentials
 }
\author{A.V. Zolotaryuk}
%\email{azolo@bitp.kiev.ua}
\address
{Bogolyubov Institute for Theoretical Physics, National Academy of
Sciences of Ukraine, Kyiv 03680, Ukraine}

\date{\today}

\begin{abstract}
Several families of one-point interactions are derived from the system consisting of 
 two and three $\delta$-potentials which are regularized by piecewise constant functions.
In physical terms such an approximating
 system represents two or three extremely thin layers separated
by some distance. 
The two-scale squeezing of this heterostructure to one point as both the width of 
 $\delta$-approximating functions and the distance between these functions
 simultaneously tend to zero
is studied using the power parameterization through a squeezing parameter $\varepsilon \to 0$,
so that the intensity of each $\delta$-potential is $c_j =a_j \varepsilon^{1-\mu}$, 
$a_j \in \R$, $j=1,2,3$,
the width of each layer $l =\varepsilon$ and the distance between the layers 
 $r = c\varepsilon^\tau$, $c >0$.
It is shown that at some values of intensities $a_1$, $a_2$ and $a_3$, 
the transmission across
the limit point interactions is non-zero, whereas outside these (resonance) values the one-point 
interactions are  opaque splitting the system at the  point of singularity  into  two 
independent subsystems. Within the interval $1 < \mu < 2$, the resonance sets consist of
two curves on the $(a_1,a_2)$-plane and three disconnected surfaces in the  $(a_1,a_2,a_3)$-space.
While approaching the parameter $\mu $ to the critical value $\mu =2$, three types of
 splitting these  sets into countable families of resonance curves and surfaces 
are observed.
  \end{abstract}


%\maketitle

%===========================================Introduction================================
\section{Introduction}
%===========================================Introduction================================

The models described by the Schr\"{o}dinger operators with singular zero-range potentials 
have widely been discussed in both the physical and mathematical literature
(see books~\cite{do1,do2,a-h,ak} for details and references). 
 These models admit exact closed analytical solutions
which describe realistic situations using different approximations via
Hamiltonians describing point interactions \cite{pc,bft,enz,c-g,ca,em}.
 Currently, because of the rapid progress in
fabricating nanoscale quantum devices, of particular importance is the point modelling of
different structures like quantum waveguides 
\cite{acf,ce}, spectral filters \cite{tc1,tc2} or infinitesimally thin sheets
 \cite{z,zz1,zz2}. A  whole body of literature (see, e.g., 
\cite{s,gh87,g,k,adk,cnp,cnt,an,n1,n2,gnn,acg,l1}, a few
to mention), including the very recent studies \cite{bn,gggm,l2,z15,kp,djp,gmmn,knt} 
with references therein, has been published  
where  the  one-dimensional  Schr\"{o}dinger operators with potentials given 
in the form of distributions are shown to 
exhibit a number of peculiar features with possible applications to quantum physics. 
A detailed list of references on this subject can also 
be found in the recent review \cite{km}.
On the other hand, using some particular regular approximations of the potential expressed
in the form of the derivative of Dirac's delta function, a number of  
interesting resonance properties of quantum particles
 tunnelling through this point potential  has been observed \cite{c-g,zci1,tn,zz14}. 
Particularly, it was found that at some values of the potential strength of the
$\delta'$-potential the transmission across this barrier is non-zero, whereas 
outside these values the barrier is fully opaque. In general terms, 
the existence of  such resonance sets in the space
of potential intensities has rigorously been established for a whole class of 
approximations of the derivative delta potential
by Golovaty with coworkers \cite{gm,gh,m1,g1,m2,gh1,g2}. This type of point interactions 
may be referred to as `resonant-tunnelling $\delta'$-potentials'. These results 
differ from those obtained within Kurasov's theory \cite{k} which was 
developed for the distributions 
defined on the space of functions discontinuous at the point of singularity.
Here the limit point interaction is also called a $\delta'$-potential. 
The common feature of Kurasov's  point potential and a resonant-tunnelling 
$\delta'$-potential is that the
transmission matrices  of both these interactions are of the diagonal form, but
the elements of these matrices are different.  It is of
interest therefore 
 to find a way where it would be possible to describe both these types in a 	
unique regularization scheme starting from the same initial regularized 
potential profile.

In the present work we address the problem on the relation between  the 
point interactions realized within Kurasov's theory and the resonant-tunnelling 
$\delta'$-potentials studied in \cite{c-g,zci1,tn,zz14,gm,gh,g1,gh1,g2,zpla10}.
We show that Kurasov's $\delta'$-potential emerges from the realistic heterostructure
 consisting of two or three extremely thin parallel plane layers separated by 
some distance. This system is studied in the limit as both the width of layers and
the distance between them simultaneously tend to zero. In such a squeezing limit,
the limit one-point interactions are proved to depend crucially on relative approaching
zero the width and the distance. As a result, different types of one-point interactions
appear in this limit depending on the way of convergence. Surprisingly, within the
same regularization scheme, it is possible to realize both the Kurasov $\delta'$-potential
and the $\kappa  \delta'$-potentials with countable sets in the $\kappa$-space 
at which a non-zero resonant tunnelling occurs. Another surprising point is that
the  $\delta$-potential discovered by \v{Seba} in \cite{s} can also be realized
together with its countable splitting at some critical point.

Consider the system consisting of $N$ parallel sheets arranged successively
   with their planes  perpendicular to the $x$-axis. The sheets are assumed 
to be homogeneous, so that one can   explore
 the one-dimensional stationary Schr\"{o}dinger equation
%----------------------------------------1--------------------------------
\begin{equation}
-\, d^2\psi(x)/dx^2 + V_{\varepsilon }(x)\psi(x) =E\psi(x)
\label{1}
\end{equation} 
%--------------------------------------------1----------------------------
where $\psi(x)$ is the wavefunction and $E$ the energy of a particle.
The potential $V_{\varepsilon }(x)$ with a squeezing parameter $\varepsilon >0$
shrinks to one point, say $x =0$, as $\varepsilon \to 0$. One of the ways to realize
limit point interactions  is to choose the potential  
$V_{\varepsilon }(x)$  in the form of a sum 
of several  Dirac's delta functions as  \cite{bn,cs,an1}
%---------------------------------------------2---------------------------------
\begin{equation}
V_\varepsilon(x) = \sum_{j =1}^N c_j(\varepsilon) \delta(x - r_j(\varepsilon)),
\label{2}
\end{equation}
%---------------------------------------------2---------------------------------
where the constants $c_j \in \R$ and the distances between the $\delta$-functions 
$r_j(\varepsilon)$  tend to zero as $\varepsilon \to 0$.
The particular case of the three-delta ($N=3$) 
spatially symmetric potential (\ref{2}), in the squeezing limit  has been studied by
Cheon and Shigehara \cite{cs}, and Albeverio and Nizhnik \cite{an1}. In this limit 
 a whole four-parameter family of point interactions has been constructed,
independently on whether or not  potential (\ref{2}) has a distributional limit. 
Here we follow the approach developed   by Exner, Neidhardt and Zagrebnov
\cite{enz}, who have approximated the $\delta$-potentials  by regular functions and 
constructed a one-point $\delta'$-interaction. In particular, they have proved that the  limit 
takes place if the distances between the `centers' of regularized potentials 
tend to zero sufficiently slow relatively to shrinking the $\delta$-approximating
 potentials. A similar 
research \cite{bft} concerns about 
 the convergence of regularized $\delta$-approximating structures to point potentials
in higher dimensions. 

In this paper we focus on the two cases when  potential  (\ref{2})
 consists of two ($N=2$) and three ($N=3$) $\delta$-potentials
separated equidistantly  by  a distance $r(\varepsilon)$ 
that  tends to zero as $\varepsilon \to 0$. 
The transmission matrices for the two- and three-delta potentials are the products
$\Lambda_{  \varepsilon} =  \Lambda_2 \Lambda_0 \Lambda_1$ and 
$\Lambda_{  \varepsilon} = \Lambda_3\Lambda_0 \Lambda_2 \Lambda_0 \Lambda_1$, 
respectively, where ($j=1,2,3$)
%--------------------------------------------3---------------------------------
\begin{eqnarray}
\!\!\!\!\!\!\!\!\!\!\!\!\!\!\!\!
  \Lambda_0  =  \left( \begin{array}{cc} ~\cos(kr)~~~~~ k^{-1}\sin(kr) \\
-\, k \sin(kr)  ~~~~~ \cos(kr) \end{array} \right) , ~~~~
\Lambda_j  =  \left( \begin{array}{cc} 1~~0 \\
c_j  ~ ~1 \end{array} \right) ,~~ k := \sqrt{E}\,.
\label{3}
\end{eqnarray}
%---------------------------------------------3-------------------------------

We restrict ourselves to
 the most simple approximation of the $\delta$-potentials by 
piecewise constant functions resulting in a three- (for $N=2$) and a  five- (for $N=3$)
layered potential profile.
In the limit as both the width of $\delta$-approximating
 functions and the distance between them tends 
to zero simultaneously we obtain different families  of one-point interactions. 
 We observe that, starting from the same profile of the three- and
five-layered structure that approximates potential (\ref{2}), 
 the limit point interactions crucially depend on the relative rate of 
 tending the width of layers and the distance between them to zero.

We follow the notations and the classification of one-point interactions
given by Brasche and Nizhnik \cite{bn}. Thus, we denote
%----------------------------------------------4---------------------------------------
\begin{equation}
\!\!\!\!\!\!\!\!\!\!\!\!\!\!\!\!\!\!\!\!\!\!\!\!\!
\left. \begin{array}{ll}
\psi_s (0) := \psi(+0)- \psi(-0),~~ &\psi'_s (0) := \psi'(+0)- \psi'(-0), \\
\psi_r (0) := \eta \psi(+0)+ (1-\eta) \psi(-0),~~ &
\psi'_r (0) := \eta \psi'(-0)+ (1-\eta) \psi'(+0),
\end{array} \right.
\label{4}
\end{equation}
%----------------------------------------------4---------------------------------------
where  $\eta \in \R$ is an arbitrary parameter (this is a  
  generalization of the generally accepted case with  $\eta =1/2$,
see, e.g., \cite{g,k,bn,l2}). Then the $\delta$-interaction, or
$\delta$-potential, with intensity $\alpha$ is defined by the boundary conditions
$\psi_s(0)=0$ and $\psi'_s(0)=\alpha \psi_r(0)$, so that
 the $\Lambda$-matrix in this case has the form
%----------------------------------------------5---------------------------------------
\begin{eqnarray}
\Lambda =  \left(
\begin{array}{cc} 1~ ~~~ 0 \\
 \alpha ~~~~ 1 \end {array} \right).
\label{5}
\end{eqnarray}
%----------------------------------------------5----------------------------------------
The dual interaction is called a $\delta'$-interaction (the notation has been suggested in 
\cite{a-h,gh87} and adopted in the literature). This point
interaction with intensity $\beta$  defined by the boundary conditions 
$\psi'_s(0)=0$ and $\psi_s(0)=\beta \psi'_r(0)$ has 
 the $\Lambda$-matrix of the form
%----------------------------------------------6---------------------------------------
\begin{eqnarray}
\Lambda =  \left(
\begin{array}{cc} 1~ ~~~ \beta \\
 0 ~~~~ 1 \end {array} \right).
\label{6}
\end{eqnarray}
%----------------------------------------------6----------------------------------------

As follows from  formulae (\ref{5}) and (\ref{6}), 
the usage of the parameter $\eta$ for both the 
$\delta$- and $\delta'$-interactions does not play any role. However, 
for the $\delta'$-potential with intensity $\gamma$ 
the potential part in equation (\ref{1}) is given by 
$\gamma \delta'(x)\psi(x)$ where the wavefunction $\psi(x)$ must be  discontinuous at $x=0$.
Therefore, due to the ambiguity of the product $\delta'(x)\psi(x)$, 
one can suppose the following generalized (asymmetric) averaging in the form
 %----------------------------------------------7-------------------------------------
\begin{equation}
\!\!\!\!\!\!\!\!\!\!\!\!\!\!\!\!\!\!\!\!\!\!\!\!\!\!\!
\delta'(x)\psi(x) = \left[ (1-\eta)\psi(-0) +\eta \psi(+0)\right]\delta'(x)+
  \left[ \eta \psi'(-0) +(1-\eta) \psi'(+0)\right] \delta(x) .
\label{7}
\end{equation}
%----------------------------------------------7----------------------------------------
This suggestion is also motivated by the studies 
 \cite{gw,vs,cnt12} which demonstrate that the plausible averaging with $\eta =1/2$
at the point of singularity in general does not work. 
The $\delta'$-potential with intensity $\gamma$ is defined by the boundary conditions
$ \psi_s(0)=\gamma \psi_r(0)$ and $\psi'_s(0)=- \gamma \psi'_r(0)$ \cite{bn}. 
An equivalent form of these conditions  is given by the $\Lambda$-matrix in the
diagonal form
%----------------------------------------------8---------------------------------------
\begin{eqnarray}
\Lambda =  \left(
\begin{array}{cc} \theta~ ~~~ 0 ~~\\
 0 ~~ ~\theta^{-1} \end {array} \right)
\label{8}
\end{eqnarray}
%----------------------------------------------8---------------------------------------
with $\theta = [ 1 +(1-\eta)\gamma ] /( 1-\eta \gamma)$.

 Finally, instead of the fourth type of point interactions 
 defined in \cite{bn} as $\delta$-magnetic potentials,
in this paper we shall be dealing with potentials which at some (resonant) 
values of intensities
are fully transparent, whereas outside these values they are completely opaque
satisfying the Dirichlet boundary conditions $\psi(\pm 0)=0$. At the resonance sets
the boundary conditions are given by one of the unit matrices
$ \Lambda = \pm \, I$, $I :=\left( \begin{array}{cc} 1~~ 0 \\
 0 ~~1 \end {array} \right)$.

%===========================================2================================
\section{A piecewise constant approximation of the $\delta$-potentials  }
%===========================================2===============================

Let us  approximate the $\delta$-potentials in (\ref{2}) by piecewise constant functions. Then
 potential (\ref{2}) is replaced by the rectangular function
%-----------------------------------------9-----------------------------
\begin{equation}
\!\!\!\!\!\!\!\!\!\!\!\!\!\!\!\!\!\!\!\!\!\!\!\!\!\!\!\!\!
V_{ lr}(x)= \left\{ \begin{array}{ll}
 0 &   \mbox{for}~~ -\infty < x < 0, ~ l < x < l+r , \\
& ~~~~~ ~2l +r <x< 2(l+r),~ 3l+2r < x< \infty , \\
 v_j     & \mbox{for}~~ (j-1)(l+r)< x < j(l+r)-r  ,~j=1,2,3,
\end{array} \right. 
\label{9}
\end{equation} 
%--------------------------------------9------------------------------
and, as a result, all the matrices 
$\Lambda_j$, $j=1,2,3$, in the product for $\Lambda_{\varepsilon }$ are replaced by 
%----------------------------------------------10-------------------------------
\begin{eqnarray}
 \Lambda_{j,l}  =  \left( \begin{array}{cc} \cos(k_j l)~~ ~~~~k_j^{-1}\sin(k_jl) \\
-\, k_j \sin(k_j l)  ~~~~~ ~~~\cos(k_j l) \end{array} \right) ,
\label{10}
\end{eqnarray}
%-----------------------------------------------10--------------------------
where 
%-------------------------------------------------11---------------------------------
\begin{equation}
 k_j : = \sqrt{  k^2 - v_j  }\, , ~~j=1,2,3.
\label{11}
\end{equation}
%-------------------------------------------------11-------------------------------
In other words, the regularized
transmission matrix $\Lambda_{lr}$ defined by the relations
%----------------------------------------12---------------------------------
\begin{eqnarray}
\!\!\!\!\!\!\!\!\!\!\!\!\!\!\!\!\!\!\!\!
\left( \begin{array}{cc} \psi(x_2)  \\
\psi'(x_2) \end{array} \right) 
 = \Lambda_{lr} \left(
\begin{array}{cc} \psi(x_1)   \\
\psi'(x_1)   \end{array} \right), ~~ \Lambda_{lr}= 
\Lambda_{3,l} \Lambda_{r}\Lambda_{2,l}\Lambda_{0}\Lambda_{1,l}  =: \left(
\begin{array}{cc} \bar{\lambda}_{11}~~ \bar{\lambda}_{12} \\
\bar{\lambda}_{21} ~~\bar{\lambda}_{22} \end{array} \right) ,
\label{12}
\end{eqnarray}
%-----------------------------------------12--------------------------------
connects the boundary conditions for the wavefunction
$\psi(x)$ and its derivative $\psi'(x)$ at $x=x_1= 0 $ and $x=x_2 =  3l+2r $
($N=3$). 
For the case of the two-delta potential ($N=2$) we set $a_3 \equiv 0$ in  potential  (\ref{9}), 
so that the boundary conditions are
 $x_1 =0$ and $x_2 = 2l +r$. The matrix elements in (\ref{12}), denoted by 
overhead bars,  depend on  the shrinking parameters $l$ and $r$, whereas in the 
limit matrix elements, if they exist, the bars are omitted, i.e., we  write
 $\lim_{ l, r \to  0 } \Lambda_{l r}  = : \Lambda =
\left( \begin{array}{cc} {\lambda}_{11}~~ {\lambda}_{12} \\
{\lambda}_{21} ~~{\lambda}_{22} \end{array} \right)$.
Having  accomplished the limit procedure,  we set $x_1 =-0$ and 
$\lim_{l,r \to 0}x_2 = +0$.

One can compute the matrix products $\Lambda_{l r} $ explicitly both for $N=2$ and
$N=3$. Using that $k_j \to \infty$,  $k_j l$ and $k_i/k_j$, $i,j=1,2,3$, 
are finite  and $r \to 0$, it is sufficient to write their truncated expressions.
As a result, we find 
the asymptotic behaviour of the elements of the matrix
$\Lambda_{ lr } =  \Lambda_{2,l} \Lambda_0 \Lambda_{1,l}$ ($N=2$):
%--------------------------------------------13-16---------------------------------------
\begin{eqnarray}
\bar{\lambda}_{11} &\to & \cos(k_1 l) \cos(k_2 l) -(k_1/k_2)\sin(k_1 l) \sin(k_2 l)
\nonumber \\
&& - k_1 r \sin(k_1 l) \cos(k_2 l), 
\label{13}\\
\bar{\lambda}_{12}& \to \,\, & 0, \label{14} \\
\bar{\lambda}_{21} &\to & -\, k_1\sin(k_1l)\cos(k_2l) -\, k_2 \cos(k_1l)\sin(k_2l)
\nonumber \\&& + k_1 k_2 r\sin(k_1l)\sin(k_2l), 
\label{15}\\
\bar{\lambda}_{22} &\to & \cos(k_1 l) \cos(k_2 l) -(k_2/k_1)\sin(k_1 l) \sin(k_2 l)
\nonumber \\ && - k_2 r \cos(k_1 l) \sin(k_2 l).
\label{16}
\end{eqnarray}
%--------------------------------------------13-16--------------------------------------
Similarly, for the three-delta potential the $\bar{\lambda}_{ij}$-asymptotes of
the matrix product $\Lambda_{ lr } = \Lambda_{3,l}\Lambda_0 
\Lambda_{2,l} \Lambda_0 \Lambda_{1,l}$ are as follows
%----------------------------------------------17-20--------------------------------
\begin{equation*}
\!\!\!\!\!\!\!\!\!\!\!\!\!\!\!\!\!\!\!\!\!\!\!\!\!\!\!\!\!\!\!\!\!\!\!
\bar{\lambda}_{11} 
 \to  \, \cos(k_1 l) \cos(k_2 l)\cos(k_3 l) -(k_1/k_2)\sin(k_1 l) \sin(k_2 l)\cos(k_3 l)
\end{equation*}
\begin{equation*}
\!\!\!\!\!\!\!\!\!\!\!\!\!\!\!\!\!\!
 -\, \, (k_1/k_3)\sin(k_1 l) \cos(k_2 l)\sin(k_3 l)
-(k_2/k_3)\cos(k_1 l) \sin(k_2 l)\sin(k_3 l)
\end{equation*}
\begin{equation*}
\!\!\!\!\!\!\!\!\!\!\!\!\!\!\!\!\!\!
 -\,\,  2k_1 r \sin(k_1 l) \cos(k_2 l)\cos(k_3 l) -k_2 r \cos(k_1 l) \sin(k_2 l)\cos(k_3 l) 
\end{equation*}
\begin{equation}
\!\!\!\!\!\!\!\!\!\!\!\!\!\!\!\!\!\!
  + \,\,  k_1 k_2 r^2 \sin(k_1 l)  \sin(k_2 l)\cos(k_3 l)
 +  (k_1 k_2 r/ k_3)\sin(k_1 l)  \sin(k_2 l)\sin(k_3 l),
\label{17} 
\end{equation}
\begin{equation}
\!\!\!\!\!\!\!\!\!\!\!\!\!\!\!\!\!\!\!\!\!\!\!\!\!\!\!\!\!\!\!\!\!\!\!
\bar{\lambda}_{12}\to \, - \, \, k_2 r^2 \cos(k_1 l) \sin(k_2 l)\cos(k_3 l)  , 
\label{18} 
\end{equation}
\begin{equation*}
\!\!\!\!\!\!\!\!\!\!\!\!\!\!\!\!\!\!\!\!\!\!\!\!\!\!\!\!\!\!\!\!\!\!\!
\bar{\lambda}_{21} \to  \, -\,\,  k_1  \sin(k_1 l) \cos(k_2 l)\cos(k_3 l)
-\, k_2 \cos(k_1 l)\sin(k_2 l)\cos(k_3 l) 
\end{equation*}
\begin{equation*}
\!\!\!\!\!\!\!\!\!\!\!\!\!\!\!\!\!\!
 -\, \, k_3 \cos(k_1 l)\cos(k_2 l)\sin(k_3 l)
+ k_1 k_2 r \sin(k_1 l)\sin(k_2 l)\cos(k_3 l) 
\end{equation*}
\begin{equation*}
\!\!\!\!\!\!\!\!\!\!\!\!\!\!\!\!\!\!
 + \, \, 2  k_1 k_3 r \sin(k_1 l)\cos(k_2 l)\sin(k_3 l) 
+ k_2 k_3 r \cos(k_1 l)\sin(k_2 l)\sin(k_3 l) 
\end{equation*}
\begin{equation*}
\!\!\!\!\!\!\!\!\!\!\!\!\!\!\!\!\!\!
+\, \,  k_1 k_3 (k_2^{-1} - k_2 r^2) \sin(k_1 l)\sin(k_2 l)\sin(k_3 l)
\end{equation*}
\begin{equation}
\!\!\!\!\!\!\!\!\!\!\!\!\!\!\!\!\!\!
 + \, \,   k^2 r^2 \cos(k_2 l) [ k_1 \sin(k_1 l)\cos(k_3 l) 
+  k_3  \cos(k_1 l)\sin(k_3 l)],
\label{19} 
\end{equation}
\begin{equation*}
\!\!\!\!\!\!\!\!\!\!\!\!\!\!\!\!\!\!\!\!\!\!\!\!\!\!\!\!\!\!\!\!\!\!\!
\bar{\lambda}_{22} \to \,  \cos(k_1 l) \cos(k_2 l)\cos(k_3 l)  
 - (k_2/k_1) \sin(k_1 l)  \sin(k_2 l)\cos(k_3 l) 
\end{equation*}
\begin{equation*}
\!\!\!\!\!\!\!\!\!\!\!\!\!\!\!\!\!\!
-\,\, (k_3/k_1) \sin(k_1 l)  \cos(k_2 l)\sin(k_3 l)
- (k_3/k_2)\cos(k_1 l) \sin(k_2 l)\sin(k_3l)  
\end{equation*}
\begin{equation*}
\!\!\!\!\!\!\!\!\!\!\!\!\!\!\!\!\!\!
 -2 k_3 r \cos(k_1 l) \cos(k_2 l)\sin(k_3l) 
- k_2 r\cos(k_1 l) \sin(k_2 l)\cos(k_3l) 
\end{equation*}
\begin{equation}
\!\!\!\!\!\!\!\!\!\!\!\!\!\!\!\!\!\!
 +\,\,  k_2k_3 r^2\cos(k_1 l) \sin(k_2 l)\sin(k_3l)
+ (k_2 k_3 r/k_1) \sin(k_1 l)  \sin(k_2 l)\sin(k_3 l).
\label{20}
\end{equation}
%------------------------------------------------17-20------------------------------------

The convergence of these elements  in the limit as both the parameters
$l$ and $r$ tend to zero will be analyzed below. Depending on the relative degree of
coming up these parameters to zero, quite different limit transmission matrices
 will be obtained realizing various one-point interactions. 

%==============================================3====================================
\section{The power parameterization of the $\Lambda_{lr}$-matrix}
%===============================================3===================================

The convergence of the transmission matrix $\Lambda_{ lr}$ as $l,r \to 0$
both for $N=2$ and $N=3$ can be parameterized through the parameter 
$\varepsilon >0$ using the positive powers $\mu $ and $\tau $  from the 
$\{ \mu >0 , \, \tau >0 \}$-quadrant as follows 
%--------------------------------------------------21----------------------------------
\begin{equation}
v_j = a_j \varepsilon^{-\mu},~~~
l=\varepsilon~~~ \mbox{and}~~~
r =c  \varepsilon^\tau,~~~c>0.
\label{21}
\end{equation}
%-------------------------------------------------21-----------------------------------
Here each coefficient $a_j \in \R$ may be called an `intensity' or a `charge' of
 the $j$th $\delta$-approximating layer.
According to (\ref{11}), we have the following asymptotic relations:
%--------------------------------------------22----------------------------------------
\begin{equation}
k_j \to  \sqrt{-\,a_j} \,
\varepsilon^{-\mu /2}, ~~
 k_jl \to  \sqrt{-\, a_j} \, \varepsilon^{1-\mu/2 }, ~~
k_j^2 l \to  -\, a_j \varepsilon^{1-\mu}.
\label{22}
\end{equation}
%--------------------------------------------22---------------------------------------
Using these relations in which  $\mu \in (0,2)$ and  $\tau \in (0, \infty)$,
 we obtain that in the limit 
as $\varepsilon \to 0$ asymptotic relations (\ref{13}), (\ref{15})-(\ref{17}), (\ref{19}) and
(\ref{20}) are reduced to
%-------------------------------------------23-24------------------------------------
\begin{eqnarray}
\bar{\lambda}_{11} &\to & \, 1+ ca_1  \varepsilon^{1-\mu +\tau} ,~~
\bar{\lambda}_{22} \to 1+  ca_2   \varepsilon^{1- \mu +\tau} ,
 \label{23} \\
\bar{\lambda}_{21} &\to & \, (a_1 + a_2)\varepsilon^{1-\mu}   +  
  ca_1 a_2 \varepsilon^{2(1-\mu )+\tau },~~~N=2;
 \label{24} 
\end{eqnarray}
%-------------------------------------------23-24-------------------------------------
%-------------------------------------------25-26---------------------------------------
\begin{equation}
\left. \begin{array}{ll}
\bar{\lambda}_{11}&  \to \, 1+ c( 2 a_1 +a_2 )\varepsilon^{ 1-\mu +\tau} +c^2 
   a_1 a_2  \varepsilon^{2(1-\mu + \tau)},\\
\bar{\lambda}_{22}& \to \,
 1+ c (  a_2  +2a_3)\varepsilon^{ 1-\mu +\tau} +c^2 
   a_2 a_3  \varepsilon^{2(1-\mu + \tau)},
\end{array} \right.
\label{25}
\end{equation}
\begin{eqnarray}
\bar{\lambda}_{21} & \to \, (a_1 + a_2  + a_3 + a_1 a_3 \varepsilon^{2-\mu})\varepsilon^{1-\mu}  + 
c(  a_1 a_2 + 2 a_1 a_3 + a_2 a_3) \varepsilon^{2(1-\mu) +\tau}  \nonumber \\
&  \,\,\,~~\,  +  c^2 a_1 a_2 a_3 \varepsilon^{3(1-\mu) +2\tau} -k^2 c^2(a_1 +a_3)
\varepsilon^{1-\mu +2\tau},~~~N=3.
 \label{26} 
\end{eqnarray}
%--------------------------------------------25-26--------------------------------------
Next, in the case 
 when $\mu =2$ and $0< \tau < \infty$,   asymptotes 
 (\ref{13}), (\ref{15}) and (\ref{16})  become 
%--------------------------------------------27-------------------------------------
\begin{equation}
\!\!\!\!\!\!\!\!\!\!\!\!\!\!\!\!\!\!\!\!\!\!\!\!\!\!\!\!\!\!\!\!\!\!\!\!\!\!
\left. \begin{array}{ll}
\bar{\lambda}_{11} \to & \left( \cos\!\sqrt{- a_1 }
 - c\sqrt{-a_1 } \sin\!\sqrt{-a_1 }\, \varepsilon^{\tau -1}\right)  \cos\!\sqrt{-a_2 }
- \sqrt{a_1/a_2 }\sin\!\sqrt{-a_1}\sin\!\sqrt{-a_2}\, ,\\
\bar{\lambda}_{22} \to & \left( \cos\!\sqrt{-a_2 }
-c \sqrt{-a_2 }  \sin\!\sqrt{-a_2 }\, \varepsilon^{\tau -1} \right)\cos\!\sqrt{-a_1 }
- \sqrt{a_2/a_1 }\sin\!\sqrt{-a_1}\sin\!\sqrt{-a_2}\, ,
\end{array} \right.
\label{27}
\end{equation}
%--------------------------------------------27----------------------------------------
%--------------------------------------------28----------------------------------------
\begin{eqnarray}
\bar{\lambda}_{21} \to & -&(\sqrt{-a_1 }\sin\!\sqrt{-a_1 }\cos\!\sqrt{-a_2 }
+\sqrt{-a_2 }\cos\!\sqrt{-a_1 }\sin\!\sqrt{-a_2 } \, )\varepsilon^{-1} \nonumber \\
& +&  c\sqrt{a_1 a_2 }\sin\!\sqrt{-a_1 }\sin\!\sqrt{-a_2 }\, \varepsilon^{\tau-2}
\label{28}
\end{eqnarray}
%--------------------------------------------28---------------------------------------
for $N=2$. Similarly, in the case with $N=3$ asymptotes (\ref{17}), (\ref{19}) and (\ref{20})
are transformed to 
%--------------------------------------------29--------------------------------
\begin{equation}
\!\!\!\!\!\!\!\!\!\!\!\!\!
\left. \begin{array}{llllllll}
\bar{\lambda}_{11}
 &\to & \left[ \cos\!\sqrt{- a_1} \cos\!\sqrt{ -a_2} \,
 -\sqrt{a_1/a_2} \sin\!\sqrt{ -a_1} \sin\!\sqrt{- a_2} \,\right. \\
&&- \, 2 c \sqrt{-a_1}   \sin\!\sqrt{ -a_1} \cos\!\sqrt{ -a_2} 
-\,c \sqrt{-a_2}  \cos\!\sqrt{ -a_1} \sin\!\sqrt{ -a_2} \, \varepsilon^{\tau -1}  \\
&& \left. +\,   c^2 \sqrt{a_1 a_2}  \sin\!\sqrt{-a_1} \sin\!\sqrt{- a_2}\, 
\varepsilon^{2(\tau -1)}\right]\cos\!\sqrt{-a_3} \\
&&  -\left[ \sqrt{-a_1}\sin\!\sqrt{- a_1} \cos\!\sqrt{ -a_2}\, 
  + \sqrt{-a_2} \cos\!\sqrt{ -a_1}\sin\!\sqrt{ -a_2} \right. \\
&& \left. -\, c \sqrt{ a_1 a_2} \sin\!\sqrt{ -a_1}  \sin\!\sqrt{-a_2}\, \varepsilon^{\tau -1} \right]
\sin\!\sqrt{- a_3}/ \sqrt{ -a_3}\, ,  \\
\bar{\lambda}_{22}
 &\to & \left[ \cos\!\sqrt{- a_2} \cos\!\sqrt{ -a_3} \,
-\sqrt{a_3/a_2} \sin\!\sqrt{ -a_2} \sin\!\sqrt{- a_3} \,\right.\\
&& -\, 2 c \sqrt{-a_3}   \cos\!\sqrt{ -a_2} \sin\!\sqrt{ -a_3} 
-\,c \sqrt{-a_2}  \sin\!\sqrt{ -a_2} \cos\!\sqrt{ -a_3} \, \varepsilon^{\tau -1}  \\
&&  + \, c^2 \sqrt{a_2 a_3}  \sin\!\sqrt{-a_2} \sin\!\sqrt{- a_3}\, 
\varepsilon^{2(\tau -1)}\,]\cos\!\sqrt{-a_1}\\
 && -\left[\sqrt{-a_2} \sin\!\sqrt{ -a_2}\cos\!\sqrt{ -a_3}\,  +
\sqrt{-a_3}\cos\!\sqrt{- a_2} \sin\!\sqrt{ -a_3} \right. \\
&&\left. -\, c \sqrt{ a_2 a_3} \sin\!\sqrt{ -a_2}  \sin\!\sqrt{-a_3}\, \varepsilon^{\tau -1}\right]
\sin\!\sqrt{- a_1}/ \sqrt{ -a_1}\, ,  \\
 \end{array} \right.
\label{29}
\end{equation}
%-----------------------------------------------29------------------------------------
%-----------------------------------------------30-------------------------------------
 \begin{equation*}
\!\!\!\!\!\!\!\!\!\!\!\!\!\!\!\!\!\!\!\!\!\!\!\!\!\!\!\!\!\!\!\!\!\!\!\!\!\!\!\!
\left. \begin{array}{lllll}
\bar{\lambda}_{21} & \to  - \, \left(
\sqrt{ -a_1}  \sin\!\sqrt{ -a_1} \cos\!\sqrt{ -a_2}\cos\!\sqrt{- a_3} 
 + \sqrt{- a_2} \cos\!\sqrt{- a_1}\sin\!\sqrt{- a_2}\cos\!\sqrt{- a_3}\right. \\ &
\left. + \sqrt{- a_3} \cos\!\sqrt{ -a_1}\cos\!\sqrt{ -a_2}\sin\!\sqrt{- a_3} \,-
\sqrt{ -a_1 a_3/a_2 } \sin\!\sqrt{- a_1}\sin\!\sqrt{ -a_2}\sin\!\sqrt{- a_3}\,
\right)\!\varepsilon^{-1}  \\
 &+ \, c\left(  \sqrt{a_1 a_2}\sin\!\sqrt{- a_1}\sin\!\sqrt{-a_2}\cos\!\sqrt{ -a_3} 
+2 \sqrt{a_1 a_3}\sin\!\sqrt{ -a_1}\cos\!\sqrt{ -a_2}\sin\!\sqrt{ -a_3} \right.\\
& + \! \left. \sqrt{a_2 a_3}\cos\!\sqrt{- a_1}\sin\!\sqrt{ -a_2}\sin\!\sqrt{- a_3}\, \right)\!
\varepsilon^{\tau -2} \\
& - \, c^2\sqrt{ -a_1 a_2 a_3} \sin\!\sqrt{ -a_1}\sin\!\sqrt{ -a_2}\sin\!\sqrt{ -a_3}\, 
\varepsilon^{2\tau -3}
\end{array} \right.
\end{equation*}
\begin{equation}
\!\!\!\!\!\!\!\!\!\!\!\!\!\!\!\!\!\!\!\!\!\!
+\,\, k^2 c^2 \! \cos\!\sqrt{ -a_2}\left( \sqrt{- a_1} \sin\!\sqrt{- a_1}\cos\!\sqrt{ -a_3}+
\sqrt{ -a_3} \cos\!\sqrt{- a_1}\sin\!\sqrt{ -a_3} \,\right)\! \varepsilon^{2\tau -1}. 
\label{30} 
\end{equation}
%---------------------------------------------30-------------------------------------

%=============================================4===========================================
\section{Admissible sets of the parameters $\mu$ and $\tau$ for realizing one-point 
interactions}
%==============================================4==========================================

\begin{figure}
\centerline{\includegraphics[width=1.0\textwidth]{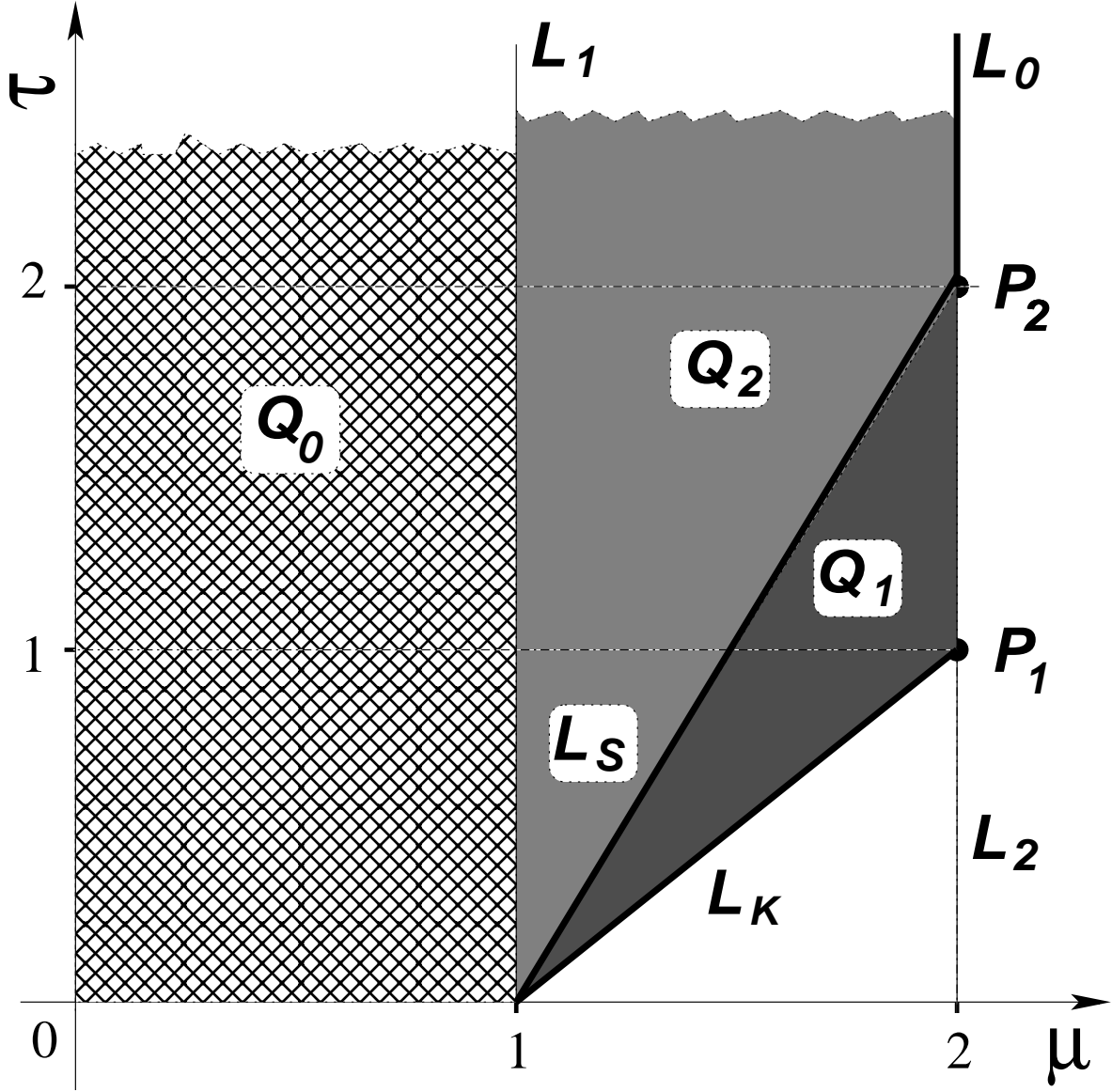}}
%\vspace{2pt}
\caption{Diagram of the  $(\mu, \tau)$-sets arranged in the correspondence with different
 families of one-point interactions to be realized from 
two- and three-delta-like layers. Notations of regions, lines and points are given
by equations (\ref{31}).
 }
\label{f1_17}
\end{figure}

For the realization of (both connected and separated)  interactions in the squeezing limit, 
 the elements $\bar{\lambda}_{11}$ and $\bar{\lambda}_{22}$ given by asymptotes
 (\ref{23}), (\ref{25}), (\ref{27})
and (\ref{29}) must be finite as $\varepsilon \to 0$. It follows from these asymptotes
that  the region  $Q :=\{0 < \mu \le 1,\, 0 < \tau < \infty \} \cup
\{1 < \mu \le 2,\, \mu -1 \le \tau < \infty \} $ is admissible for the finiteness of
the limit elements $\lambda_{11}$ and $\lambda_{22}$. 
Concerning the asymptote for $\bar{\lambda}_{12}$ in the case with $N=3$, we have in the 
region $Q$ the limits 
$\bar{\lambda}_{12} \to c^2 a_2 \varepsilon^{1-\mu +2\tau } \to 0 $ for $0< \mu < 2$
and $\bar{\lambda}_{12} 
\to -\, c^2 \sqrt{-a_2}\cos\!\sqrt{-a_1} \sin\!\sqrt{-a_2}\cos\!\sqrt{-a_3}\, 
\varepsilon^{2\tau -1} \to 0$ for $\mu =2$. Therefore the convergence of the 
matrix $\Lambda_{lr}$ in the limit as $\varepsilon \to 0$ has to be analyzed in 
the region $Q$ shadowed in figure~\ref{f1_17}. 
While the matrix element $\bar{\lambda}_{12}$ for both 
$N=2$ and $N=3$ has the zero limit, 
 the elements $\bar{\lambda}_{21}$ given by asymptotes (\ref{24}), (\ref{26}),
(\ref{28}) and (\ref{30})
are in general divergent as $\varepsilon \to 0$. In the region $Q$, the highest 
divergence is of the order $\varepsilon^{1-\mu}$, $1< \mu \le 2$. 
However, on some sets of $Q$ and 
at some conditions  on the intensities $a_j$, the coefficients at 
the divergent term $\varepsilon^{1-\mu}$ can be zero. These conditions will be analyzed
below in detail, but now we single out those sets of $Q$ where this analysis will
be carried out. Thus,  we split the region $Q$
into the sets as shown in figure~\ref{f1_17}  
%------------------------f1----------------------------
%-------------------------f1------------------------------------- 
using the following definitions:
%--------------------------------------------------31------------------------------
\begin{eqnarray}
& \mbox{region}~ &Q_0 :=\{ 0 < \mu < 1,\, 0< \tau < \infty \}; \nonumber \\
& \mbox{lines}& L_0 := \{ \mu = 2,\, 2 < \tau < \infty \},\nonumber \\
&&  L_1:= \{ \mu = 1,\, 0< \tau < \infty \},~~
L_2 := \{ \mu = 2,\, 0 < \tau < \infty \}, ~~~\nonumber \\
&&  L_K := \{ 1< \mu < 2,\, \tau =\mu-1 \},~
 L_S := \{ 1< \mu < 2,\, \tau =2(\mu-1) \}; ~~~\nonumber \\
&\mbox{regions}~& Q_1 := \{ 1< \mu < 2,\, \mu-1 <\tau < 2(\mu-1) \},~~\nonumber \\
&& Q_2 := \{ 1< \mu < 2,\, 2(\mu -1) <\tau < \infty \}; \nonumber \\
&\mbox{points}~& P_1 := \{ \mu =2, \, \tau = 1 \},~~P_2 := \{ \mu = \tau = 2 \}.
\label{31}
\end{eqnarray}
%--------------------------------------------------31------------------------------
In the region $Q_0$, the transmission is perfect because each layer is fully transparent
in the squeezing limit, so that this point interaction is trivial. At the boundary 
of this region, i.e., on the line $L_1$, the squeezing limit of each layer
leads to the $\delta$-potential with the total intensity  equal 
 the algebraic sum of the layer intensities, i.e., $\alpha = a_1 +a_2 +a_3$
in $\Lambda$-matrix (\ref{5}). 

There exists a possibility to consider the line $\tau = 3(\mu - 1)/2$, 
$1< \mu < 2$, if both the equations 
$a_1 +a_2 +a_3 =0$ and $a_1 a_2 +2a_1 a_3 + a_2 a_3 =0$ are fulfilled simultaneously. Excluding
$a_3$ from these equations, we find the condition $a_1^2 +(a_1 +a_2)^2 =0$ which is valid only if
$a_1 =a_2 =0$ and therefore $a_3 =0$. Similarly, 
we have  to analyze the case $\tau =3/2$ at $\mu =2$ in (\ref{30}). Here
 the $\varepsilon \to 0$ limit of $\bar{\lambda}_{21}$ will be finite 
if both the coefficients at $\varepsilon^{-1}$
and $\varepsilon^{\tau -2}$  equal zero  simultaneously resulting in 
two equations. Excluding  from these equations the term
$\sqrt{ a_3}\tan\!\sqrt{ a_3}$, we find the condition 
$ a_1 \tan^2\!\sqrt{ a_1}\cos^2\!\sqrt{ a_2} + \left( \sqrt{ a_1 }\tan\!\sqrt{ a_1}
+\sqrt{ a_2}\tan\!\sqrt{ a_2} \right)^2 =0 $ which can also be satisfied  
if  $a_1= a_2= a_3=0 $ and therefore the case with $\tau =3/2$
does not produce connected point interactions for $1 < \mu \le 2$.

Finally, as follows from asymptotes (\ref{27})-(\ref{30}),  on the whole line $L_2$, 
there are the point subsets of intensities for which $\sin\!\sqrt{-a_j}=0$, $j=1,2$ ($N=2$) and 
$j=1,2,3$ ($N=3$). For this case 
we have $\Lambda =\pm \, I$ and therefore the limit point interactions are reflectionless.
 They are `non-interacting' wells because in the case of a single well ($N=1$) 
the full transmission across this well occurs if $\sin\!\sqrt{-a_1} =0$. Thus, so far we have 
examined the one-point interactions with full transmission (non-resonant in
the region $Q_0$ and resonant on the line $L_2$) and non-resonant $\delta$-potentials
realizing on the line $L_1$.

%=======================================================5==================================
\section{One-point interactions at the line $L_K$}
%========================================================5=================================

\begin{figure}
\centerline{\includegraphics[width=1.0\textwidth]{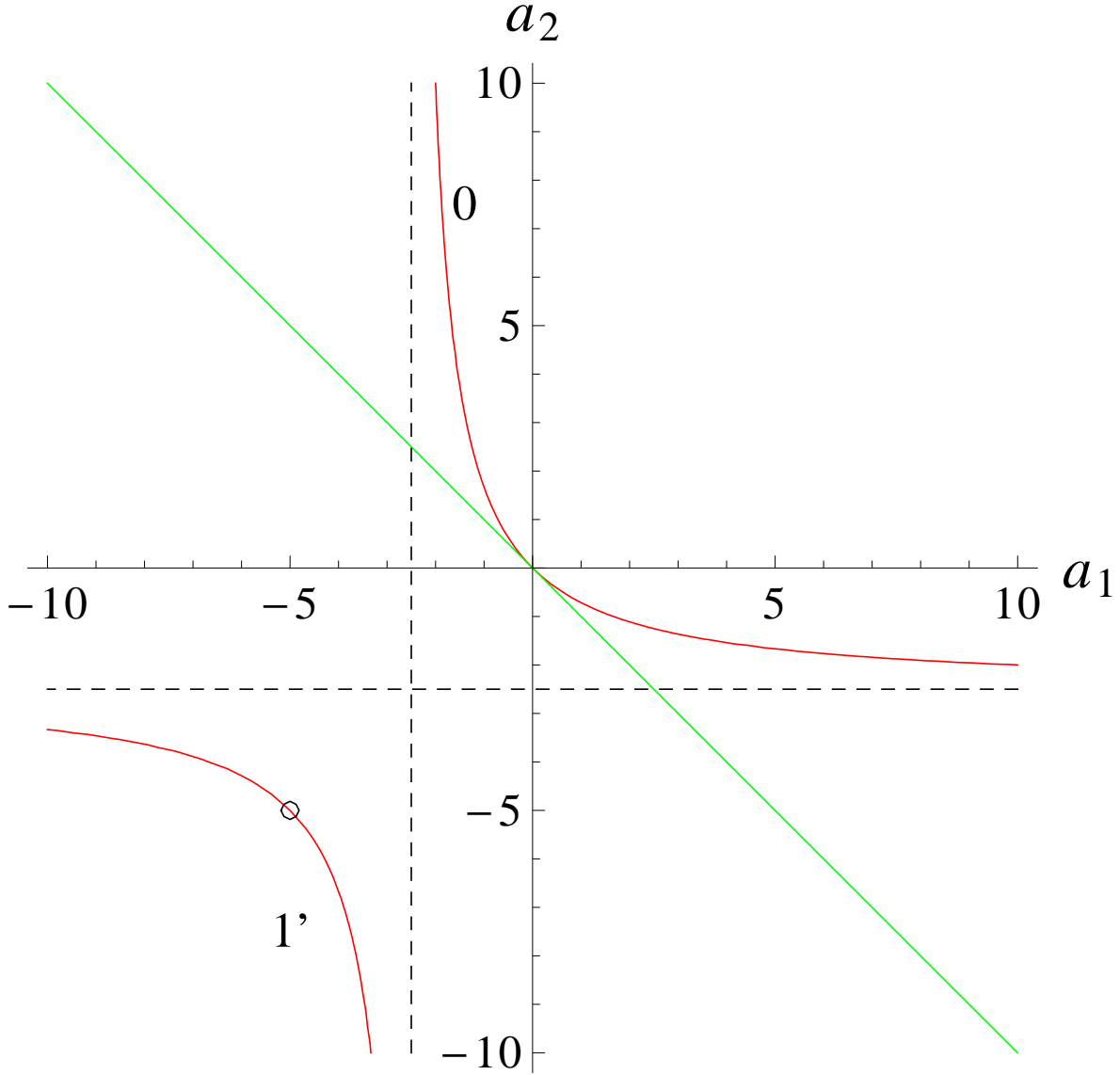}}
%\vspace{2pt}
\caption{
Two disconnected resonance curves (0 and 1', red lines) as a solution of 
the first equation (\ref{32})
 with $c=0.4$ forming the ${\cal K}_2$-set. Curve 0  corresponds to the 
two barrier-well configurations
($a_1 a_2 <0$), while curve 1' describes the  resonance related to the double-well structure.
The point at line 1' with the coordinates $a_1=a_2= -\, 2/c$ corresponds to the 
symmetric double-well system with full transmission ($\lambda =-\, I$). The line 
$a_1 +a_2=0$ (green, also called  the ${\cal L}_2$-set)
 intersects the resonance curve 0 only at the origin 
$a_1=a_2 =0$. The coordinates of the asymptotic (dashed) lines are $a_1 =-c^{-1}$ and  
$a_2 =-c^{-1}$. When $c \to 0$, the zeroth resonance curve 0 remains to be `pinned' 
to the origin $a_1 =a_2 =0$  straightening to the ${\cal L}_2$-line, while 
the resonance curve 1' vanishes `escaping' to infinity.
}
\label{f2_17}
\end{figure}

\begin{figure}
\centerline{\includegraphics[width=1.0\textwidth]{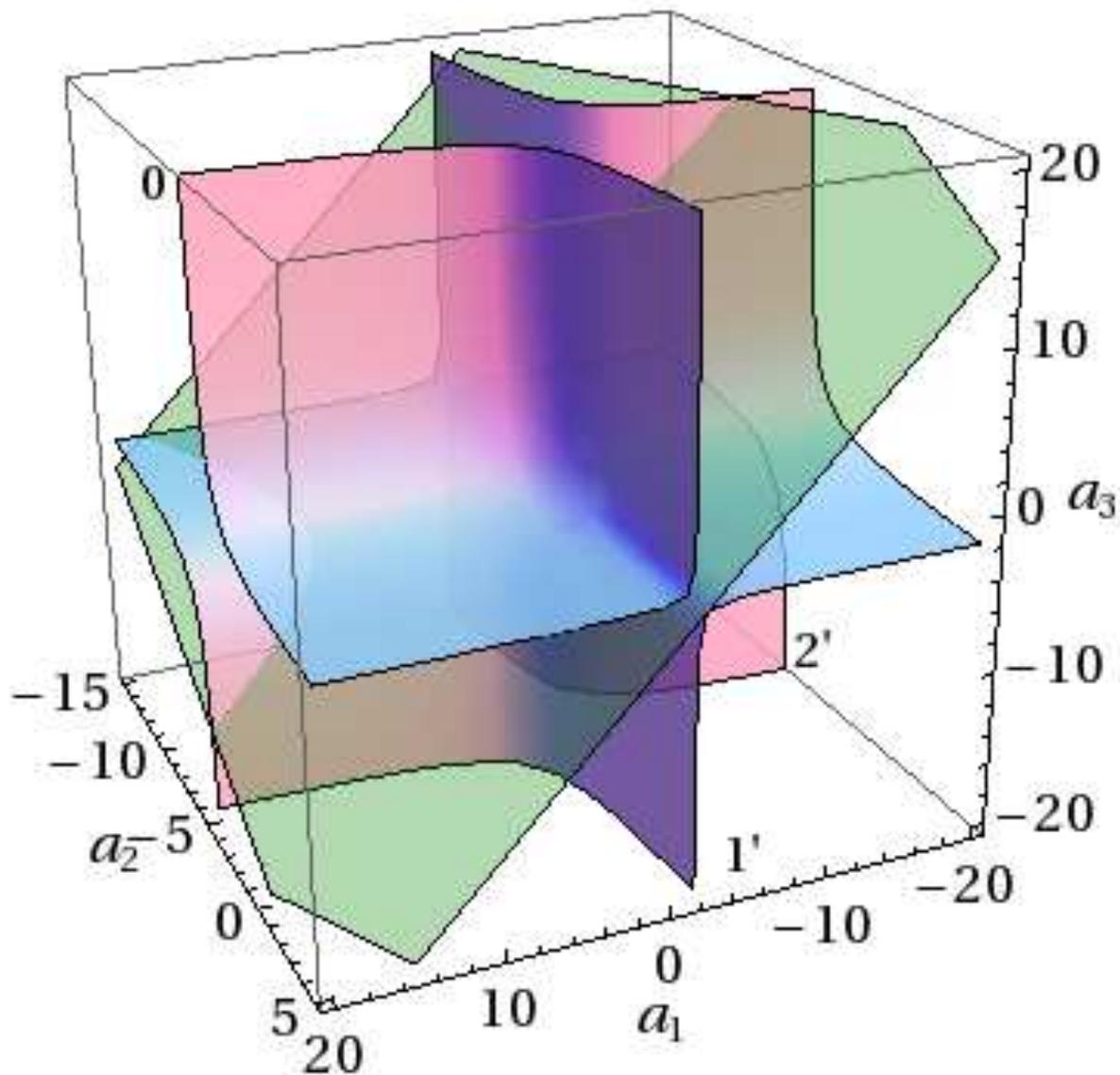}}
%\vspace{2pt}
\caption{Three disconnected resonance surfaces (0, 1' and 2') as a solution 
of the second equation (\ref{32}) with $c=0.4$ forming the ${\cal K}_3$-set. 
The zeroth  surface 0
corresponds to the six barrier-well configurations (one or two wells are present, i.e.,
$a_1 >0 $, $a_2 < 0$, $a_3 >0$;  $a_1 >0 $, $a_2 <0$, $a_3 <0$;
$a_1 >0 $, $a_2 > 0$, $a_3 <0$; $a_1 <0 $, $a_2 < 0$, $a_3 >0$; $a_1 <0 $, $a_2 > 0$,
 $a_3 <0$; $a_1 < 0 $, $a_2 > 0$, $a_3 >0$). This surface does not lie in  the octants 
$a_1 >0 $, $a_2 > 0$, $a_3 >0$ and $a_1 <0 $, $a_2 < 0$, $a_3 <0$. It is `pinned' to 
the origin $a_1 =a_2=a_3=0$  straightening to the plane  $a_1 +a_2+a_3=0$ 
(green, also called the ${\cal L}_3$-set) as $c \to 0$,  
while the two other surfaces 1' and 2' vanish `escaping' to infinity. 
The ${\cal L}_3$-plane   intersects 
only with the zeroth  surface 0 forming the two disconnected curves being a solution
of equation (\ref{40}), on which the
standard distributional limit $V_\varepsilon(x) \to \kappa \delta'(x)$ does exist.
 }
\label{f3_17}
\end{figure}

One of the ways to remove the divergence of the element $\bar{\lambda}_{21}$ in  (\ref{26})	
 as $\varepsilon \to 0$ is the requirement that the total group of terms at the divergent term
$\varepsilon^{1-\mu}$, $1< \mu < 2$, has to be zero, i.e., 
$\lim_{\varepsilon \to 0}
\left(\varepsilon^{\mu-1}\bar{\lambda}_{21}\right)_{\tau =\mu-1}=0$. 
Indeed, this requirement can be satisfied on the line $L_K$ if
the conditions 
%-----------------------------------------------32------------------------------
\begin{equation}
\!\!\!\!\!\!\!\!\!\!\!\!\!\!\!\!\!\!\!\!\!\!\!\!\!\!\!\!\!\!\!\!\!\!\!\!\!\!\!\!
\left. \begin{array}{ll}
 ~~~~ K_2(a_1,a_2; c) : = a_1 +a_2 +ca_1a_2 =0 & \mbox{for}~ N=2,  \\
  K_3 (a_1,a_2,a_3; c) : =  a_1 +a_2 + a_3 +c(a_1 a_2 +2  a_1 a_3 +
  a_2 a_3) + c^2a_1 a_2 a_3 =0 & \mbox{for} ~ N=3, 
\end{array} \right.
 \label{32}
\end{equation}
%------------------------------------------32---------------------------------
are fulfilled. At fixed coefficient $c$, these conditions may be considered as 
equations with respect to  the intensities $a_1$, $a_2$ and $a_3$. 
The first of these equations ($N=2$) has been 
obtained earlier by Brasche and Nizhnik \cite{bn}. In what follows we shall also be dealing 
with other conditions like (\ref{32}). They may be referred to as `resonance equations'
and their solutions `resonance sets'. The solutions of equations
(\ref{32}) are plotted in figures~\ref{f2_17} and \ref{f3_17}, respectively for
$N=2$ and 3. In each of these two cases, the curve and surface marked
 in these figures by 0 appear to be  `pinned' to the origins
$a_1=a_2=0$ and $a_1=a_2=a_3=0$. They are considered as  background 
 branches of the corresponding resonance sets and therefore we call them the `zeroth resonance'
curve ($N=2$, figure~\ref{f2_17}) and surface ($N=3$, figure~\ref{f3_17}). In the limit
as $c \to 0$, the curve 1' in figure~\ref{f2_17} and the surfaces 1' and 2' in
figure~\ref{f3_17} vanish `escaping' to infinity. At the same time, the zeroth branches 0
straighten to the line $a_1 +a_2 =0$ and the plane  $a_1 +a_2 +a_3=0$, respectively. 
%---------------------------------------------fig2-------
%-------------------------------------------------fig2------------- 
%------------------------------------------------fig3----
%--------------------------------------------------fig3------------ 

Using  next equations (\ref{32}) in asymptotes (\ref{23}) and (\ref{25}), we obtain 
the diagonal elements of the limit $\Lambda$-matrix:
%--------------------------------------------33-----------------------------------
\begin{equation}
\left. \begin{array}{ll}
\lim_{\varepsilon \to 0}\bar{\lambda}_{11}  =:
\theta = \left\{ \begin{array}{ll}
 1+ca_1 &   \mbox{for}~~  N=2, \\
1 +c( 2a_1 + a_2) +c^2 a_1  a_2  & \mbox{for} ~~ N=3,
\end{array} \right.  \\
\lim_{\varepsilon \to 0}\bar{\lambda}_{22} 
 =: \rho = \left\{ \begin{array}{ll}
 1+c a_2 &   \mbox{for}~~  N=2, \\
1 +c( a_2 + 2a_3)  +c^2 a_2  a_3  & \mbox{for} ~~ N=3.
\end{array} \right. \end{array} \right.
\label{33}
\end{equation}
%--------------------------------------------33----------------------------------
In virtue of equations (\ref{32}), we have $\rho =\theta^{-1}$ and therefore on the 
$L_K$-line 
 the limit transmission matrix becomes of diagonal form (\ref{8}).
Setting here  $\theta = (2+\gamma)/(2-\gamma)$ and 
$\rho = (2-\gamma)/(2+\gamma)$,
we obtain Kurasov's $\delta'$-potential with intensity $\gamma \in \R \setminus 
\{ \pm 2\}$ defined in the distributional sense 
on the space of discontinuous at $x=0$ test functions. Next, we find the 
resonance values of $a_1$, $a_2$ and $a_3$ as functions of the strength $\gamma$:
%---------------------------------------------34--------------------------------------
\begin{equation}
a_1 =  {2\gamma \over c( 2- \gamma) },~~
a_2 = - {2\gamma \over c(2+  \gamma)}
\label{34}
\end{equation}
%----------------------------------------------34-------------------------------------
for $N=2$ and 
%------------------------------------------------35--------------------------------------
\begin{equation}
a_1 = {1 \over c( 2+ca_2)} \left( { 2\gamma \over 2- \gamma  } -ca_2 \right), ~~
a_3 = -{1 \over c( 2+ca_2)} \left(  { 2\gamma \over 2+ \gamma  } +ca_2 \right), 
\label{35}
\end{equation}
%---------------------------------------------------35---------------------------------------
with arbitrary $a_2 \in \R \setminus \{-a_2/c \}$,  for $N=3$.
In particular, for $N=2$  the barrier-well structure corresponds to the interval
$-2 < \gamma < 2$ ($a_1 >0,~a_2<0$ for $-2 < \gamma <0$ and $a_1 <0,~a_2>0$ 
for $0 < \gamma < 2$), whereas beyond this interval ($2< |\gamma|< \infty$), we have
 the double-well configuration.  We call the vectors $(a_1,a_2) \in \R^2$ and  
$(a_1,a_2,a_3) \in \R^3 $ that satisfy equations
(\ref{32}) the resonance sets ${\cal K}_2$ and ${\cal K}_3$, respectively.
 Next, the point interactions realized on the line $L_K$  are referred in the 
following to as `resonant-tunnelling $\delta'$-potentials of the ${\cal K}$-type',
despite that the double-well case is involved here as well,
 together with a barrier-well structure. 

Finally, it should be noticed that there exists a particular  subfamily  of
the intensities $a_1$ $a_2$ and $a_3$ from the ${\cal K}_{2,3}$-sets 
 for which $\theta = \pm \, 1$ in (\ref{8}), realizing
 the point interactions with full transmission.  Thus, for $N=2$ these values are
$a_1 =a_2 =-2/c$ (this point is indicated in figure~\ref{f2_17})
 resulting in the  matrices $\Lambda =- \, I$. 
In the three-delta case the two conditions $a_1 =a_3$ and $2a_1 +a_2 +ca_1a_2  = 0$ 
provide the matrix $\Lambda =I$, whereas the other two conditions
$a_1 +a_3 =-2/c$ (in general, an asymmetric structure) and $2a_1 +a_2+ca_1a_2 =-2/c$
 lead to the matrix $\Lambda =-\,I$.

%===============================================6======================================
\section{\v{S}eba's transition at the line $L_S$}
%==============================================6=========================================

Other types of one-point interactions can be realized in the regions $Q_1$ and $Q_2$
including the line $L_S$ that separates these regions (see figure~\ref{f1_17}).  
In fact, as shown below, the point interactions on these three sets are related to
those studied by \v{S}eba in \cite{s} (see Theorem 3 therein). First, we note that
in the limit as $\varepsilon \to 0$,
 the divergence of the elements $\bar{\lambda}_{21}$ given by (\ref{24}) and (\ref{26})	
can be excluded at the line $L_S$ and in the region $Q_2$ if the following 
 (resonance) conditions
%--------------------------------------------36----------------------------------
\begin{equation}
\!\!\!\!\!\!\!\!\!\!\!\!\!\!\!\!\!\!\!\!\!
\left. \begin{array}{ll}
   ~~~~K_2(a_1,a_2; c)\vert_{c=0}=: L_2(a_1,a_2)  =  a_1 +a_2 =0 ~~ & \mbox{for}~~N=2,  \\
 K_2(a_1,a_2,a_3 ; c)\vert_{c=0}=:  L_3(a_1,a_2,a_3)  =   a_1 +a_2 + a_3=0 ~~ & 
\mbox{for}~~N=3
\end{array} \right.
\label{36}
\end{equation}
%--------------------------------------------36-----------------------------------
hold true, being just a `linearized' version of equations (\ref{32}). In what follows
we refer the line  $a_1+a_2=0 $ (green in figure~\ref{f2_17}) and the plane 
$a_1 +a_2 +a_3=0 $ (figure~\ref{f3_17}) to as ${\cal L}_2$- and ${\cal L}_3$-sets,
respectively. Next, in virtue of
(\ref{23}) and (\ref{25}), we have  $\bar{\lambda}_{11}, \, \bar{\lambda}_{22} \to 1$,
whereas $\bar{\lambda}_{21} \to I$ in the region $Q_2$ and $\bar{\lambda}_{21} \to \alpha$
with
%-----------------------------------------37-----------------------------
\begin{equation}
\alpha= -\, c \left\{ \begin{array}{ll}
   a_1^2 =a_2^2  &   \mbox{for} ~N=2, \\
a_1^2 +(a_1 + a_2)^2= a_1^2 +a_3^2 =  (a_2 + a_3)^2 + a_3^2  & \mbox{for}~N=3
\end{array} \right. 
\label{37}
\end{equation} 
%--------------------------------------37-------------------------
at the line $L_S$. Below this line, in the region $Q_1$, the $\bar{\lambda}_{21}$-term
is divergent as $\varepsilon \to 0$. Since the limits of   $\bar{\lambda}_{11}$ and 
$\bar{\lambda}_{22}$ are finite, we get  in this region
 the  separated interactions 
satisfying the  Dirichlet conditions $\psi(\pm 0)=0$.

Thus, the point interactions realized in the region $Q_1 \cup Q_2 \cup L_S$ exhibit  
   the transition of transmission that occurs on the  resonance ${\cal L}_{2,3}$-sets while
varying the rate of increasing the distance $r$ between the $\delta$-approximating potentials
in (\ref{9}).
For sufficiently slow squeezing  [$\mu -1< \tau < 2(\mu -1)$] this distance, the limit
point interactions  are opaque, for intermediate shrinking 
[$\tau =2(\mu -1)$] the interactions become  partially 
transparent ($\delta$-well) and for fast shrinking [$2(\mu -1) < \tau < \infty $]
the interactions appear to be fully   transparent. In other words, 
the line $L_S$ separates the regions $Q_1$ of full reflection  and $Q_2$ of
perfect transmission.

Finally, note  that, contrary to the  point interactions of the $\delta$-potential type 
realized on the line $L_1$ where the intensities $a_1$  and $a_2$ ($N=2$) or
$a_1$, $a_2$ and $a_3$ ($N=3$) are arbitrary and the intensity of the total
$\delta$-potential is just the algebraic sum $a_1 +a_2$ or $a_1 +a_2 +a_3$, 
on the line $L_S$,
 the non-zero transmission across the limit $\delta$-well potential,
but now with intensity (\ref{37}),
 occurs only on the resonance line and plane $L_{2,3}=0$ referred in the following 
to as ${\cal L}_{2,3}$-sets. Similarly, the interactions realized at the line $L_S$
are called `resonant-tunnelling $\delta$-potentials  of the ${\cal L}$-type' and ones  
in the region $Q_2$ `resonant-tunnelling  reflectionless
potentials of the ${\cal L}$-type'.

Let us consider now potential (\ref{2}) with $N=3$  rewritten in the form 
%---------------------------------------------38---------------------------------
\begin{equation}
\!\!\!\!\!\!\!\!\!\!\!\!\!\!\!\!\!\!\!
V_\epsilon(x) = (c/\epsilon)^{\sigma}\left[ a_1 \delta(x) +a_2\delta(x-\epsilon)+
a_3 \delta(x-2\epsilon)\right],~~0< \sigma \le 1,
\label{38}
\end{equation}
%---------------------------------------------38---------------------------------
with a new squeezing parameter  $\epsilon>0$. For the case with $N=2$
[$a_3 \equiv 0 $ in (\ref{38})], \v{S}eba has proved (see Theorem 3 in  \cite{s}) 
that at $\sigma =1/2$ the limit point interaction is the $\delta$-potential 
described by the $\Lambda$-matrix of form (\ref{5}) with
the intensity $\alpha $ given by the first equation (\ref{37})  if $a_1 +a_2 =0$,
being in fact a resonance condition for intensities $a_1$ and $a_2$. 
At this condition for all $\sigma \in (0, 1/2)$ the limit $\Lambda$-matrix is the unit,
while for $\sigma \in (1/2, 1)$ the limit point interactions are separated
satisfying the two-sided Dirichlet conditions $\psi(\pm 0)=0$.
In physical terms, the value $\sigma =1/2$ is a `transition' point (at which the
transmission is partial) separating the opaque potentials from
those with full transmission.

In fact, \v{S}eba's theorem can be extended to the case with $N=3$. Indeed,
by a straightforward calculation of the matrix product  
 $\Lambda_{  \epsilon} = \Lambda_3\Lambda_0 \Lambda_2 \Lambda_0 \Lambda_1$
with matrices (\ref{3}) in which $c_j = a_j(c/\epsilon)^{\sigma}$ and $r=\epsilon$, 
we find the limits $\bar{\lambda}_{11}, \, \bar{\lambda}_{22} \to 1$ and 
$\bar{\lambda}_{12} \to 0$ for $\sigma \in (0,1)$. The singular element 
$\bar{\lambda}_{21}$ has the asymptote
%----------------------------------------------39---------------------------------
\begin{equation}
\!\!\!\!\!\!\!\!\!\!\!\!\!\!\!\!\!\!\!\!\!\!\!\!\!\!\!\!\!
\bar{\lambda}_{21} \to (a_1 +a_2 +a_3)(c/\epsilon)^{\sigma} 
+(a_1 a_2 +2a_1a_3 +a_2a_3)\epsilon(c/\epsilon)^{2\sigma }+
 a_1 a_2 a_3 \epsilon^2(c/\epsilon)^{3\sigma}
\label{39}
\end{equation}
%-----------------------------------------------39------------------------------
as $\epsilon \to 0$. It follows from this asymptote that under the resonance 
condition $a_1 + a_2 + a_3=0$ the limit $\delta $-potential is realized at
$\sigma =1/2$ with the  intensity $\alpha $ given by the second formula (\ref{37}). 

The value $\sigma = 2/3$ which could result in a finite $\epsilon \to 0$ limit of 
$\bar{\lambda}_{21}$ is not admissible because  the equations 
$a_1 + a_2 + a_3=0$ and $a_1 a_2 +2a_1a_3 +a_2a_3=0$ can be fulfilled simultaneously
 only if $a_1 =a_2 =a_3 = 0$. However, as follows from (\ref{39}), at $\sigma =1$
we have $\bar{\lambda}_{21} \to 0$ if $(a_1,a_2,a_3) \in {\cal K}_3$. 
Moreover, in this case the $\epsilon \to 0$ limits of $\bar{\lambda}_{11}  $ and 
$\bar{\lambda}_{22}  $ are given by the second formulae (\ref{33}). 

All these results obtained above for potential (\ref{38}) appear to be in agreement
with those obtained for both the lines  $L_K$ and $L_S$. Indeed, the comparison of
expression  (\ref{38}) with potential (\ref{2}) in which $N=2,3$ leads to the equations
$c_j = a_j \varepsilon^{1- \mu}$ and $r= c \varepsilon^\tau$. As a result,
we find the relation $\tau = \sigma^{-1}(\mu-1) $, so that
at $\sigma =1$ we have the resonance sets defined by equations (\ref{32}) on the line 
$L_K$, while at $\sigma=1/2$, i.e., on the line $L_S$, we find that 
 formulae (\ref{37}) hold true. 

Thus, all the interactions realized at $\sigma =1$ or the same on the line $L_K$
are of Kurasov's $\delta'$-potential type defined via the test functions which are
discontinuous at $x=0$. It is of interest the problem whether or not the standard
$\delta'$-potentials defined on the $C_0^\infty$ functions can be found among
the ${\cal K}$-families defined by conditions (\ref{32}). This problem has been 
examined by Brasche and Nizhnik \cite{bn} including also the case when the fourth  
$\delta$-potential, i.e., the term  $a_4\delta(x-3\epsilon)$ 
 is added in the square brackets of (\ref{38}). In our notations these results read 
as follows. For $N=2$, in virtue of the first equation (\ref{32}), the distributional 
limit $V_\epsilon(x) \to \kappa \delta'(x)$ cannot exist because it is necessary
that $a_1 +a_2 =0$. However, for $N=3$ and 4, such a limit can be realized
on the planes $a_1 + a_2 +a_3 =0$ and  $a_1 + a_2 +a_3 +a_4=0$, respectively.
For instance, in the case with $N=3$ we have $\kappa = c(a_1 -a_3)$ where $a_1$ and
$a_2$ are connected through the quadratic equation
%--------------------------------------------40----------------------------------
\begin{equation}
 (1+ca_1)a_3^2 + (1+ca_3)a_1^2 =0.
\label{40}
\end{equation}
%---------------------------------------------40-------------------------------------
Solving the last equation with respect to $a_1$ or $a_3$, one can be convinced
that there exists a non-empty set of solutions to this equation. 
The existence of these solutions is illustrated in figure~\ref{f3_17} as the intersection
of  the  resonance surface 1'   with the ${\cal L}_3$-plane.

%=============================================7=====================================
\section{Splitting of resonance sets  at the critical point $ \mu =2$ 
  } 
%=============================================7===================================

\begin{figure}
\centerline{\includegraphics[width=1.0\textwidth]{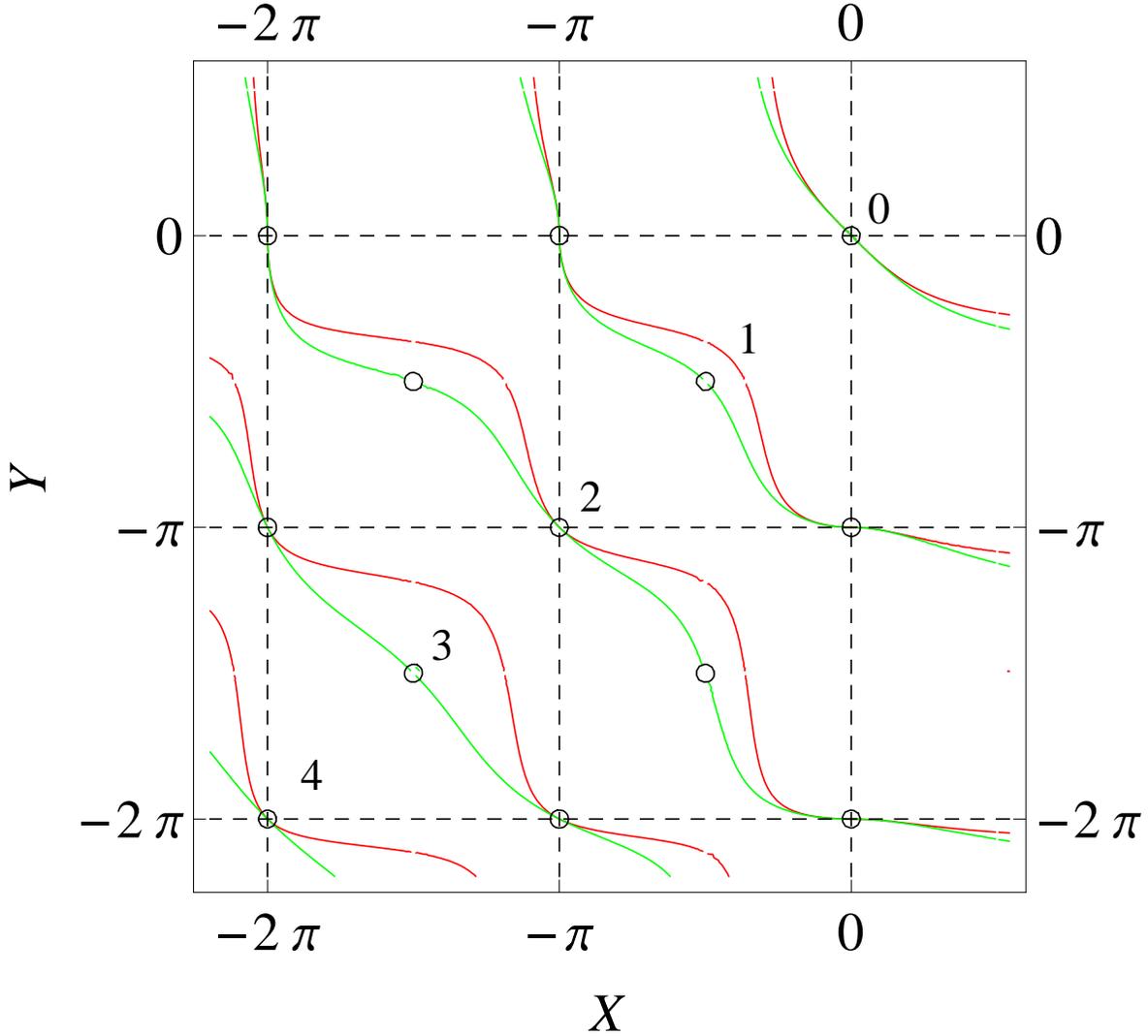}}
%\vspace{2pt}
\caption{ The first five ($n=0,1,2,3,4$) resonance curves as solutions of 
equation (\ref{41}) at  $c=0.4$ (red) and $c=0$ (green) forming the ${\cal F}_2$- and 
${\cal J}_2$-sets, respectively. While approaching 
the critical value $\mu =2$, the zeroth  curve 0 in figure~\ref{f2_17} is deformed 
 remaining to be `pinned' to the origin $a_1 =a_2=0$, while the curve 1' 
vanishes. Instead of the latter curve, the 
 countable set of  curves numbered by $n=1,2,\ldots $  detaches from the curve 0
in  the negative directions of $X$-  and $Y$-axes involving in addition double-well configurations. 
At the points which are indicated with empty balls, the full transmission across
the configurations without barriers takes place ($\Lambda = \pm \, I$).  
 }
\label{f4_17}
\end{figure}

Consider now the realization of point interactions at the limiting right-hand points
of the lines $L_K$ and $L_S$ ($P_1$ and $P_2$, respectively), and the limiting right-hand
line ($L_0$) of the region $Q_2$ (see figure 1). Thus, at $\tau =1$, 
the total coefficients at the divergent term in the singular matrix elements
$\bar{\lambda}_{21}$ given by (\ref{28}) and (\ref{30}) become zero if
the  resonance equations
%--------------------------------------------41--------------------------------------
\begin{eqnarray}
 F_2(a_1, a_2;c) && :=  
 \sqrt{-a_1 }\, \sin\!\sqrt{-a_1 }\, \cos\!\sqrt{-a_2 }\,  \nonumber \\
&& \,\, + \left(\cos\!\sqrt{-a_1 } -c  \sqrt{-a_1 } \, \sin\!\sqrt{-a_1 }\, 
\, \right)\!\sqrt{-a_2 }\,
\sin\!\sqrt{-a_2 } =0
 \label{41}
\end{eqnarray}
%--------------------------------------------41---------------------------------------
for $N=2$ and 
%----------------------------------------------42----------------------------------
\begin{equation*}
\!\!\!\!\!\!\!\!\!\!\!\!\!\!\!\!\!\!\!\!\!\!\!\!\!\!\!\!\!\!\!\!\!
F_3(a_1, a_2, a_3;c)  : = [\sqrt{-a_1}\sin\!\sqrt{-a_1 }
\cos\!\sqrt{-a_2 }\, +\sqrt{-a_2}\cos\!\sqrt{-a_1 }\sin\!\sqrt{-a_2 } 
\end{equation*}
\begin{eqnarray}
&& - c \sqrt{a_1 a_2 } \sin\!\sqrt{- a_1}\sin\!\sqrt{ -a_2}\,\, ]\cos\!\sqrt{-a_3 }
 + [\cos\!\sqrt{-a_1 }\cos\!\sqrt{-a_2 }\,
\nonumber \\
&& -2c\sqrt{-a_1}\sin\!\sqrt{-a_1 }\cos\!\sqrt{-a_2 }\, 
-c \sqrt{-a_2}\cos\!\sqrt{-a_1 }\sin\!\sqrt{-a_2 }  
\nonumber \\
&& +\left( c^2 \sqrt{a_1 a_2}  -\sqrt{a_1/a_2}\, \right)
\sin\!\sqrt{-a_1 }\sin\!\sqrt{-a_2 }\,\,] \sqrt{-a_3 }\sin\!\sqrt{-a_3 }=0
\label{42}
\end{eqnarray}
%----------------------------------------------42----------------------------------
for $N=3$ hold true. 
Note that $F_2(a_1,a_2)=F_2(a_2,a_1)$ and $F_3(a_1,a_2, a_3) = F_3(a_3,a_2, a_1)$.
The solutions of the equation $F_2(a_1,a_2;c)=0$  for two values of the coefficient $c$ 
are plotted in figure~\ref{f4_17}. In order to display in the figure the maximum 
number of resonance curves, instead of  $a_1$ and $a_2$,  the new rescaled coordinates
$X := \mbox{sign}(a_1)|a_1|^{1/2}$ and $Y := \mbox{sign}(a_2)|a_2|^{1/2}$
have been introduced.  
%------------------------fig4----------------------------
%-------------------------fig4------------------------------------ 
%------------------------fig5----------------------------
\begin{figure}
\centerline{\includegraphics[width=1.0\textwidth]{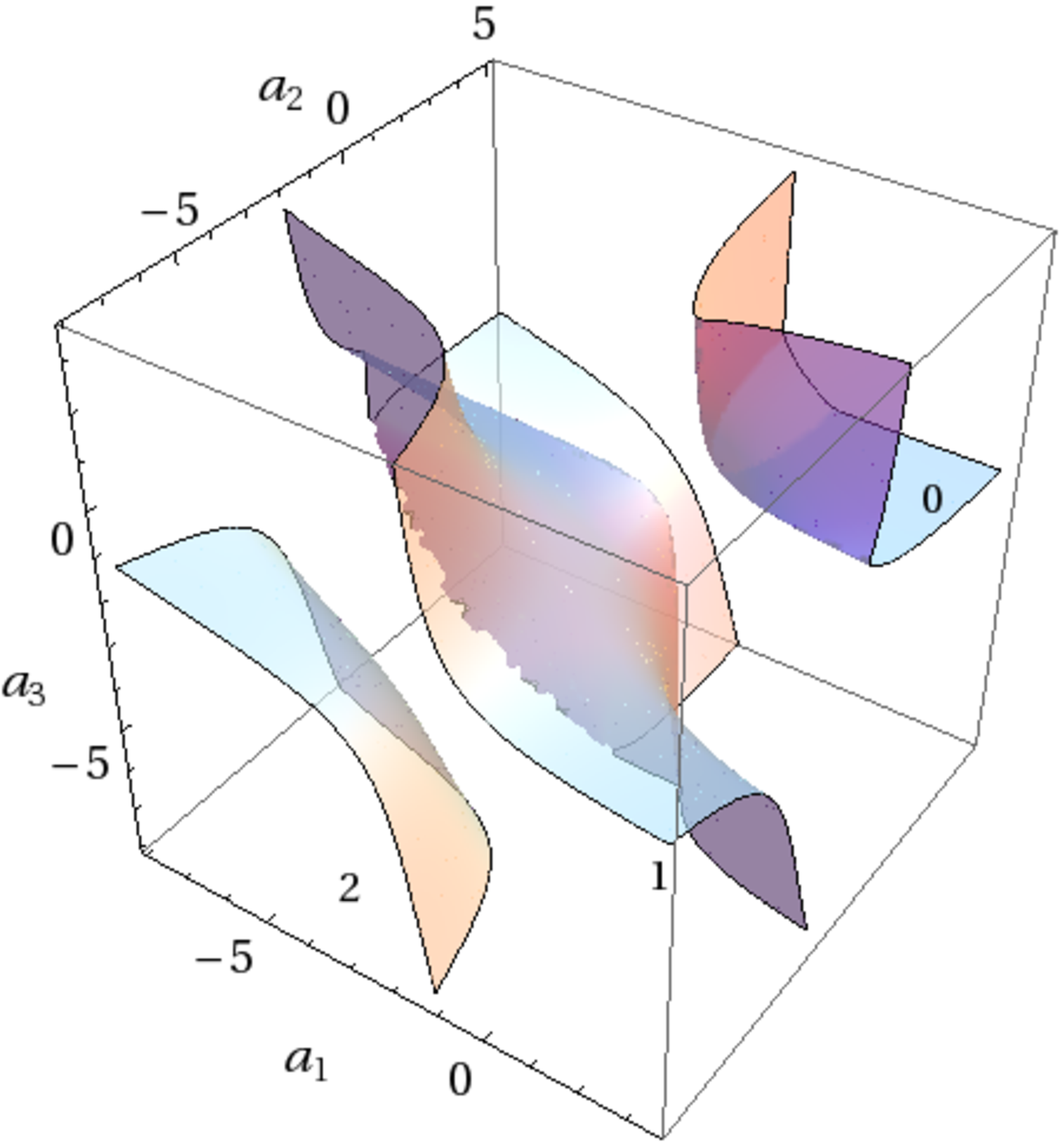}}
%\vspace{2pt}
\caption{ The first three ($n=0,1,2$) resonance surfaces as solutions of 
equation (\ref{42}) with $c=0$. While approaching 
the critical value $\mu =2$, the zeroth resonance surface 0 
in figure~\ref{f3_17} is deformed 
 remaining to be `pinned' to the origin $a_1 =a_2=a_3=0$,
while  surfaces  1' and 2' vanish. Instead of the latter  surfaces, the 
 countable set of  surfaces numbered by $n=1,2, \ldots $  detaches from  the surface 0
 involving in addition triple-well configurations. 
 }
\label{f5_17}
\end{figure}
%-------------------------f5------------------------------------- 

For the sake of brevity of the formulae that will appear in the further analysis 
 of the  point interactions to be realized on the limiting sets
$P_1$, $P_2$ and $L_0$,  we introduce the following functions:
%--------------------------------------------43---------------------------------
\begin{equation}
\!\!\!\!\!\!\!\!\!\!\!\!\!\!\!\!\!\!\!\!\!\!\!\!\!\!\!\!\!\!\!\!\!
f_2(a_1,a_2;c):=  { \cos\!\sqrt{-a_1} -c \sqrt{-a_1}\sin\!\sqrt{-a_1} \over
 \cos\!\sqrt{-a_2}} \, , ~~~g_2(a_1,a_2):= -\,{ \sqrt{-a_1}\sin\!\sqrt{-a_1}
\over  \sqrt{-a_2}\sin\!\sqrt{-a_2} }\,;
\label{43}
\end{equation}
%---------------------------------------------43-----------------------------------
%-----------------------------------------------44------------------------------
\begin{equation}
\!\!\!\!\!\!\!\!\!\!\!\!\!\!\!\!\!\!\!\!\!\!\!\!\!\!\!\!\!\!\!\!\!\!
\left. \begin{array}{ll}
f_3(a_1,a_2,a_3;c) := \left[\,\cos\!\sqrt{-a_1 }\cos\!\sqrt{-a_2 } 
-2 c\sqrt{-a_1}\sin\!\sqrt{-a_1 } \cos\!\sqrt{-a_2 } \right. \\
~~~~~~~~~~~~~~~~~~~~~~~  - \,c\sqrt{-a_2}\cos\!\sqrt{-a_1 }\sin\!\sqrt{-a_2 } \\
~~~~~~~~~~~~~~~~~~~~~~~  + \left. \left( c^2 \sqrt{a_1a_2}-\sqrt{a_1/a_2} \, \right)
 \sin\!\sqrt{-a_1 }\sin\!\sqrt{-a_2 }   \, \right]/ 
\cos\!\sqrt{-a_3 }\, , \\
g_3(a_1,a_2,a_3;c):=- \left(\sqrt{- a_1} \sin\!\sqrt{ -a_1}\cos\!\sqrt{- a_2}\, +
\sqrt{- a_2} \cos\!\sqrt{ -a_1}\sin\!\sqrt{- a_2} \right. \\
~~~~~~~~~~~~~~~~~~~~~~~ \left.
 -\, c \sqrt{a_1 a_2 } \sin\!\sqrt{- a_1}\sin\!\sqrt{ -a_2}\, 
\right)/\sqrt{- a_3} \sin\!\sqrt{- a_3}  
\end{array} \right.
 \label{44}
\end{equation}
%------------------------------------------44---------------------------------
for any $c \ge 0$. By straightforward calculations one can prove that, under resonance 
conditions (\ref{41}) and 
(\ref{42}), $f_2(a_1,a_2;c)=g_{2}(a_1,a_2)$ and  $f_3(a_1,a_2,a_3;c)=
g_{3}(a_1,a_2,a_3;c)$ as well as 
%-------------------------------------------45------------------------------------------
\begin{equation}
\!\!\!\!\!\!\!\!\!\!
 f_2(a_2,a_1;c)=f_2^{-1}(a_1,a_2;c)~~~\mbox{and}~~~f_3(a_3,a_2,a_1;c)
=f_3^{-1}(a_1,a_2,a_3;c).
\label{45}
\end{equation}
%-------------------------------------------45-----------------------------------------

Thus, in virtue of properties  (\ref{45}),
 the family of point interactions realized at the point $P_1$ is described 
by the $\Lambda$-matrix of type (\ref{8}) with 
%-----------------------------------------46----------------------------
\begin{equation}
\theta= \left\{ \begin{array}{ll}
   f_2(a_1,a_2;c) =g_2(a_1,a_2) &   \mbox{for} ~N=2, \\
 f_2(a_1,a_2,a_3;c) =g_2(a_1,a_2,a_3;c) & \mbox{for}~N=3,
\end{array} \right. 
\label{46}
\end{equation} 
%--------------------------------------46------------------------
where  the intensities $a_j$, $j=1,2,3$,
 satisfy the resonance equations (\ref{41}) and (\ref{42}), respectively.
 
The solutions of  transcendental equations (\ref{41}) and (\ref{42}) 
determine the countable 
sets of resonance curves on the $(a_1,a_2)$-plane and resonance surfaces
in the $(a_1,a_2,a_3)$-space which are plotted in figures~\ref{f4_17} and \ref{f5_17},
respectively. 
We refer these resonance curves and surfaces
to as ${\cal F}_2$- and ${\cal F}_3$-sets, respectively.
The limit transmission matrix on these sets is of 
diagonal form (\ref{8}) with the element $\theta$
 given by equations (\ref{44})-(\ref{46}). 
The  point interactions of this countable family 
may be called  `multiple-resonant-tunnelling
$\delta'$-potentials of the ${\cal F}$-type'.

At the point $P_2$, with $\tau =2$ in  (\ref{28}) and (\ref{30}), we find
that    these divergent terms become finite 
in the $\varepsilon \to 0$ limit  under  conditions (\ref{41}) and (\ref{42}) in
which $c$ is formally set zero, i.e., if the equations
%--------------------------------------------47---------------------------------------
\begin{equation}
\!\!\!\!\!\!\!\!\!\!\!\!\!\!\!\!\!\!\!\!\!\!\!\!\!\!\!\!\!
 J_2(a_1,a_2) : = F_2(a_1,a_2;c)\vert_{c=0}=0~~\mbox{and}~~ 
J_3(a_1,a_2,a_3) : = F_3(a_1,a_2,a_3;c)\vert_{c=0}=0 
\label{47}
\end{equation}
%---------------------------------------------47--------------------------------
are fulfilled. Similarly to the ${\cal F}_{2,3}$-sets, the solutions of these equations
yield  curves on the $(a_1,a_2)$-plane (see green curves in figure~\ref{f4_17}) 
and surfaces in the $(a_1,a_2,a_3)$-space 
that may be referred to as ${\cal J}_2$- and ${\cal J}_3$-sets, respectively.
Under  resonance conditions (\ref{47}), the $\varepsilon \to 0$ limits of 
(\ref{28}) and (\ref{30}) become finite producing a family of one-point
interactions described by the transmission matrix of the type
%----------------------------------------------48---------------------------------------
\begin{eqnarray}
\Lambda =  \left(
\begin{array}{cc} \theta~ ~~~ 0 ~~\\
 \alpha ~~ ~\theta^{-1} \end {array} \right),
\label{48}
\end{eqnarray}
%----------------------------------------------48---------------------------------------
where 
%----------------------------------------------49--------------------------------------
\begin{equation}
\!\!\!\!\!\!\!\!\!\!\!\!\!\!\!\!\!\!\!\!\!\!
\alpha= c\left\{ \begin{array}{lll} 
\sqrt{a_1 a_2 }\sin\!\sqrt{-a_1 }\sin\!\sqrt{-a_2 } \, , & N=2, \\
\sqrt{a_1 a_2}\sin\!\sqrt{ -a_1}\sin\!\sqrt{ -a_2}\cos\!\sqrt{- a_3} \\
~~~~~~~~~~~~~~~~~~~~~
+ 2 \sqrt{a_1 a_3}\sin\!\sqrt{- a_1}\cos\!\sqrt{- a_2}\sin\!\sqrt{- a_3} & \\
~~~~~~~~~~~~~~~~~~~~~
+\sqrt{a_2 a_3}\cos\!\sqrt{- a_1}\sin\!\sqrt{ -a_2}\sin\!\sqrt{ -a_3}\, , & N=3 .
\end{array} \right.
\label{49}
\end{equation}
%-----------------------------------------------49------------------------------------
and the diagonal elements are given by formulae (\ref{46}) in which again $c$ is
formally set zero, i.e., 
%-----------------------------------------50---------------------------
\begin{equation}
\theta= \left\{ \begin{array}{ll}
   f_2(a_1,a_2;c)\vert_{c=0} =g_2(a_1,a_2) &   \mbox{for} ~N=2, \\
 f_2(a_1,a_2,a_3;c)\vert_{c=0} =g_2(a_1,a_2,a_3;c)\vert_{c=0} & \mbox{for}~N=3.
\end{array} \right. 
\label{50}
\end{equation} 
%--------------------------------------50------------------------

The elements $\theta$ and $\alpha$ in $\Lambda$-matrix (\ref{48}) 
are determined by the countable sets of solutions
to resonance equations (\ref{47}). Therefore,  this family of one-point interactions
 may be called `resonant-tunnelling ($\delta' + \delta$)-potentials of
the ${\cal J}$-type'. 

Finally, for $\tau >2$ (at the line $L_0$) the divergent terms in (\ref{28}) and (\ref{30})
vanish at all at the same resonance conditions (\ref{47}).
Next, from expressions 
(\ref{27}) and (\ref{29}) we conclude that the limit $\Lambda$-matrix is of form (\ref{8})
with the element $\theta$ defined by formulae (\ref{50}).
Similarly,  this family of point interactions
 may be called `resonant-tunnelling $\delta'$-potentials of
the ${\cal J}$-type'. For $N=2$ this type of one-point interactions coincides
with that examined  earlier in  \cite{c-g,gm}.

Thus, we have defined in this section the resonance
${\cal F}_{2}$-,  ${\cal F}_{3}$- and  ${\cal J}_{2,3}$-sets as solutions of transcendental
equations (\ref{41}), (\ref{42}) and (\ref{47}), respectively.   These sets consist of
the infinite number of curves (for $N=2$) and surfaces (for $N=3$) which are numbered
by $0,1,2, \dots$. On the other hand, in  the open sets  $L_K$, $L_S$ and $Q_2$,
we have defined above 
  the resonance ${\cal K}_{2,3}$- and  ${\cal L}_{2,3}$-sets as solutions of
equations (\ref{32}) and (\ref{36}), respectively, consisting of two curves ($N=2$,
numbered by 0 and 1') and three surfaces ($N=3$, numbered by 0, 1' and 2'). 
The  phenomenon of `splitting' these resonance sets is observed 
while approaching the right-hand limiting sets: the point $P_1$ and the  line $P_2 \cup L_0$.
The mechanism of this splitting can be explained as follows. The curve 1' in
figure~\ref{f2_17} and the surfaces 1' and 2' in figure~\ref{f3_17}, which are 
`unpinned' to the  origins $a_1 =a_2 =0$ and  $a_1 =a_2 =a_3 =0$,  vanish  
 while approaching the  limiting sets. At the same time, the zeroth (background)
resonance subsets marked in figures~\ref{f4_17} and \ref{f5_17} by 0 remain to be
`pinned' to the origins, modifying their shape and becoming to be embedded in the 
angles with the vertices at $a_1 = a_2 = -\, b_0 $ where $b_0$ is a solution 
of the equation $\cot\!\sqrt{b}= c\sqrt{b}$ and   $a_1 =a_2 =a_3= -(\pi/2)^2$,
respectively for $N=2$ and 3. Instead of the curve 1' and the surfaces
1' and 2', the detachment of the resonances numbered by $n=1,2, \ldots$  from
the  zeroth resonances 0 occurs forming countable sets. Thus, figure~\ref{f4_17} illustrates
the resonance curves with $n=0,1,2,3,4$ for the two cases with $c>0$ (${\cal F}_2$-set) 
and $c=0$ (${\cal J}_2$-set). In figure~\ref{f5_17} the three resonance  surfaces with
 $n=0,1,2$ from the ${\cal J}_3$-set are plotted. 

To conclude this section, 
 it should be noticed  that everywhere beyond 
all the resonance sets  described above on  the sets $L_0$,  $L_K$, $L_S$,
$P_1$, $P_2$ and $Q_2$, the point interactions are separated, similarly to
the region $Q_1$, where they satisfy  the Dirichlet conditions $\psi(\pm 0)=0$ 
and act as a fully reflecting wall.

%==============================================8========================================
\section{Concluding remarks}
%============================================8=============================================

In this paper, the system consisting of two or three
$\delta$-potentials (with intensities $a_j \in \R$, $j=1,2$ if $N=2$ and 
$j=1,2,3$ if $N=3$) has been approximated in the most simple way, namely
by piecewise constant functions and then the convergence of
the corresponding transmission matrices has been studied in the squeezing limit
as both the width of  $\delta$-approximating functions $l$ and the
distance between them $r$ tend to zero. 
The admissible rates of 
shrinking the parameters $l$ and $r$ have been controlled through power parameterization
 (\ref{21}), involving the two powers 
$\mu$ and $\tau$ as well as the squeezing parameter $\varepsilon \to 0$. For convenience of
the presentation, we have used the diagram of admissible values for $\mu$ and $\tau$
plotted in figure~\ref{f1_17}.
Using this parameterization as well as the piecewise constant approximation of 
the $\delta$-functions in potential (\ref{2}) with $N=2,3$, 
it was possible to get the explicit
expressions  for the corresponding $\Lambda$-matrices and to
 treat thus the reflection-transmission properties of the one-point
interactions directly.

Starting from the  same  three-layer (for $N=2$)
and   five-layer (for $N=3$) potential profile given by (\ref{9}), 
a whole family of  limit one-point interactions 
with resonant-tunnelling behaviour has been realized.  For the interactions  
 realized at the line $L_K$, the resonance sets referred to as ${\cal K}_{2,3}$ 
 consist of two curves on the $(a_1,a_2)$-plane  ($N=2$) and three surfaces in the 
$(a_1,a_2,a_3)$-space  ($N=3$).  In its turn, for the interactions obtained 
in the region $L_S \cup Q_2$, the resonance  sets called ${\cal L}_{2,3}$ appear to be
 the line $a_1 +a_2 =0$
($N=2$) and the plane $a_1 + a_2 + a_3=0$ ($N=3$).
While approaching the right-hand limiting sets   
($L_K \to P_1$, $L_S \to P_2$ and $Q_2 \to L_0$), the ${\cal K}$- 
and ${\cal L}$-sets split into infinite but countable sets resulting in
the set transformations ${\cal K}_{2,3} \to {\cal F}_{2,3}$ and ${\cal L}_{2,3}
\to {\cal J}_{2,3}$. These transformations are illustrated by the figures
from \ref{f2_17} to \ref{f5_17}. In other words, at the limiting sets, 
the detachment of the countable ${\cal F}$- and ${\cal J}$-sets from 
the ${\cal K}$- and ${\cal L}$-sets takes place, respectively. 
Accordingly,  the  point interactions realized on these sets belong to
 ${\cal K}$-, ${\cal L}$-, ${\cal F}$- and ${\cal J}$-families and their $\Lambda$-matrices
are given by (\ref{8}) with (\ref{33}), (\ref{5}) with (\ref{37}),  (\ref{8}) with 
(\ref{46}), and (\ref{48}) with (\ref{49}) and (\ref{50}) as well as (\ref{8}) with (\ref{50}).
Outside all these resonance sets as well as for all $a_j$, $j=1,2,3$, in the region 
$Q_1$,   the point interactions are separated with
the Dirichlet boundary conditions $\psi(\pm 0)=0$.

 In principle, a similar straightforward 
analysis could be carried out for higher $N$ resulting in 
the same types of one-point interactions with  
resonance sets  ${\cal K}_N$, ${\cal L}_N$, ${\cal F}_N$ and ${\cal J}_N$
being $(N-1)$-dimensional hypersurfaces,
however, the corresponding formulae appear to be quite complicated. 
 To conclude, it should be noticed that the approach developed in this paper can 
 be a starting point for further studies on regular approximations of point 
interactions and understanding  the resonance mechanism.

\bigskip
{\bf  Acknowledgments}
\bigskip

The author acknowledges the financial support from the Department of Physics and Astronomy
of the National Academy of Sciences of Ukraine under project No.~0117U000240.
 He is indebted to one of Referees for criticisms and recommendations,  
resulting in the significant improvement of the paper. 
 He would like to express gratitude to 
Yaroslav Zolotaryuk for stimulating discussions and valuable suggestions.

\end{document}